\documentclass{cimento}
\usepackage{graphicx}
\usepackage{amssymb}
\usepackage{amsmath}

\newcommand{\be}{\begin{equation}}
\newcommand{\ee}{\end{equation}}
\newcommand{\bea}{\begin{eqnarray}}
\newcommand{\eea}{\end{eqnarray}}

\def\kpn{K^+\rightarrow\pi^+\nu\bar\nu}
\def\klpn{K_{L}\rightarrow\pi^0\nu\bar\nu}

\newcommand{\mev}{\, {\rm MeV}}

\newcommand{\vcb}{|V_{cb}|}
\newcommand{\vtd}{|V_{td}|}
\newcommand{\vub}{|V_{ub}|}
\newcommand{\vts}{|V_{ts}|}

\title{Hints for new sources of  flavour violation
in meson mixing}
\author{M.~Blanke}
\instlist{\item
Institut f\"ur Kernphysik, Karlsruhe Institute of Technology,
  Hermann-von-Helmholtz-Platz 1,\\
  D-76344 Eggenstein-Leopoldshafen, Germany
\item Institut f\"ur Theoretische Teilchenphysik,
  Karlsruhe Institute of Technology, Engesserstra\ss e 7,\\
  D-76128 Karlsruhe, Germany}

\PACSes{\PACSit{11.30.Hv}{Flavor symmetries}
\PACSit{14.40.Df}{Strange mesons}
\PACSit{14.40.Nd}{Bottom mesons}}

\begin{document}

\maketitle

\begin{abstract}
The recent results by the Fermilab Lattice and MILC collaborations on the hadronic matrix elements entering $B_{d,s}-\bar B_{d,s}$ mixing show a significant tension of the measured values of the mass differences $\Delta M_{d,s}$ with their SM predictions. We review the implications of these results in the context of Constrained Minimal Flavour Violation models. In these models, the CKM elements $\gamma$ and $|V_{ub}|/|V_{cb}|$ can be determined from $B_{d,s}-\bar B_{d,s}$ mixing observables, yielding a prediction for $\gamma$ below its tree-level value. Determining subsequently $|V_{cb}|$ from the measured value of either $\Delta M_s$ or $\varepsilon_K$ gives inconsistent results, with the tension being smallest in the Standard Model limit. This tension can be resolved if the flavour universality of new contributions to $\Delta F = 2$ observables is broken. We briefly discuss the case of $U(2)^3$ flavour models as an illustrative example.
\end{abstract}

\section{Introduction}

Despite the impressive experimental and theoretical progress that has recently been made both at the high energy and the precision frontier, no clear sign of new physics (NP) has emerged so far. Yet, some intriguing anomalies exist, in particular in the flavour sector \cite{Aaij:2014pli,Aaij:2014ora,Huschle:2015rga,Aaij:2015yra,Buras:2015yba}. Precise predictions of flavour changing neutral current (FCNC) observables in the Standard Model (SM) are therefore of utmost importance.

Recently the Fermilab Lattice and MILC collaborations (Fermilab-MILC) presented new and improved results for the hadronic matrix elements \cite{Bazavov:2016nty}
\be\label{Kronfeld}
 F_{B_s}\sqrt{\hat B_{B_s}}=(276.0\pm8.5)\mev,\qquad  F_{B_d} \sqrt{\hat B_{B_d}}=
(229.4\pm 9.3)\mev 
\ee
entering the mass differences in $B_s-\bar B_s$ and $B_d - \bar B_d$ mixing, respectively, as well as their ratio
\be\label{xi}
\xi=\frac{F_{B_s}\sqrt{\hat B_{B_s}}}{F_{B_d}\sqrt{\hat B_{B_d}}}=1.203\pm0.019\,.
\ee
Using these results together with the CKM matrix elements determined from global fits, the authors of \cite{Bazavov:2016nty} identified tensions between the measured values of $\Delta M_s$, $\Delta M_d$ and their ratio and their SM predictions by $2.1\sigma$, $1.3\sigma$ and $2.9\sigma$, respectively.

As discussed in \cite{Blanke:2014asa}, relating deviations from the SM in different meson system allows to shed light on the NP flavour structure. The simplest flavour structure an NP model can have is the one of the SM. {\it I.\,e.}\ flavour is broken only by the SM Yukawa couplings $Y_u$ and $Y_d$ and no new effective operators are present beyond the SM ones. This scenario is known as Constrained Minimal Flavour Violation (CMFV) \cite{Buras:2000dm,Buras:2003jf,Blanke:2006ig}. An important consequence of this hypothesis is the flavour universality on NP contributions to FCNC processes in the various meson systems. All new contributions to meson mixing observables can therefore be described by a single real and flavour-universal function
\be\label{eq:S}
S(v) = S_0(x_t) + \Delta S(v)\,,
\ee
where $v$ collectively denotes the new parameters in a given model. Here $S_0(x_t)$ is the SM one-loop function, and the new CMFV contribution $\Delta S(v)$ has been shown to be non-negative \cite{Blanke:2006yh}.

In \cite{Blanke:2016bhf} the consequences of the Fermilab-MILC results \cite{Bazavov:2016nty} for CMFV models have been analysed in detail. We summarise the findings of \cite{Blanke:2016bhf} in what follows.

\section{Universal unitarity triangle 2016}

\begin{figure}[!h]
\centering{\includegraphics[width=6.5cm]{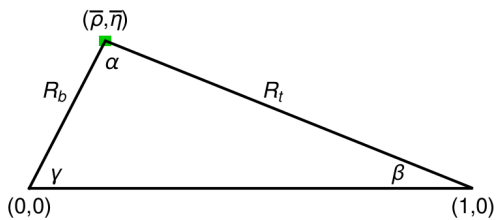}  }   
\caption{Universal unitarity triangle as of February 2016. The uncertainty on the apex $(\bar\rho,\bar\eta)$ is depicted by the green rectangle. From \cite{Blanke:2016bhf}.\label{fig01}}
\end{figure}

\begin{figure}
\includegraphics[width=6.5cm]{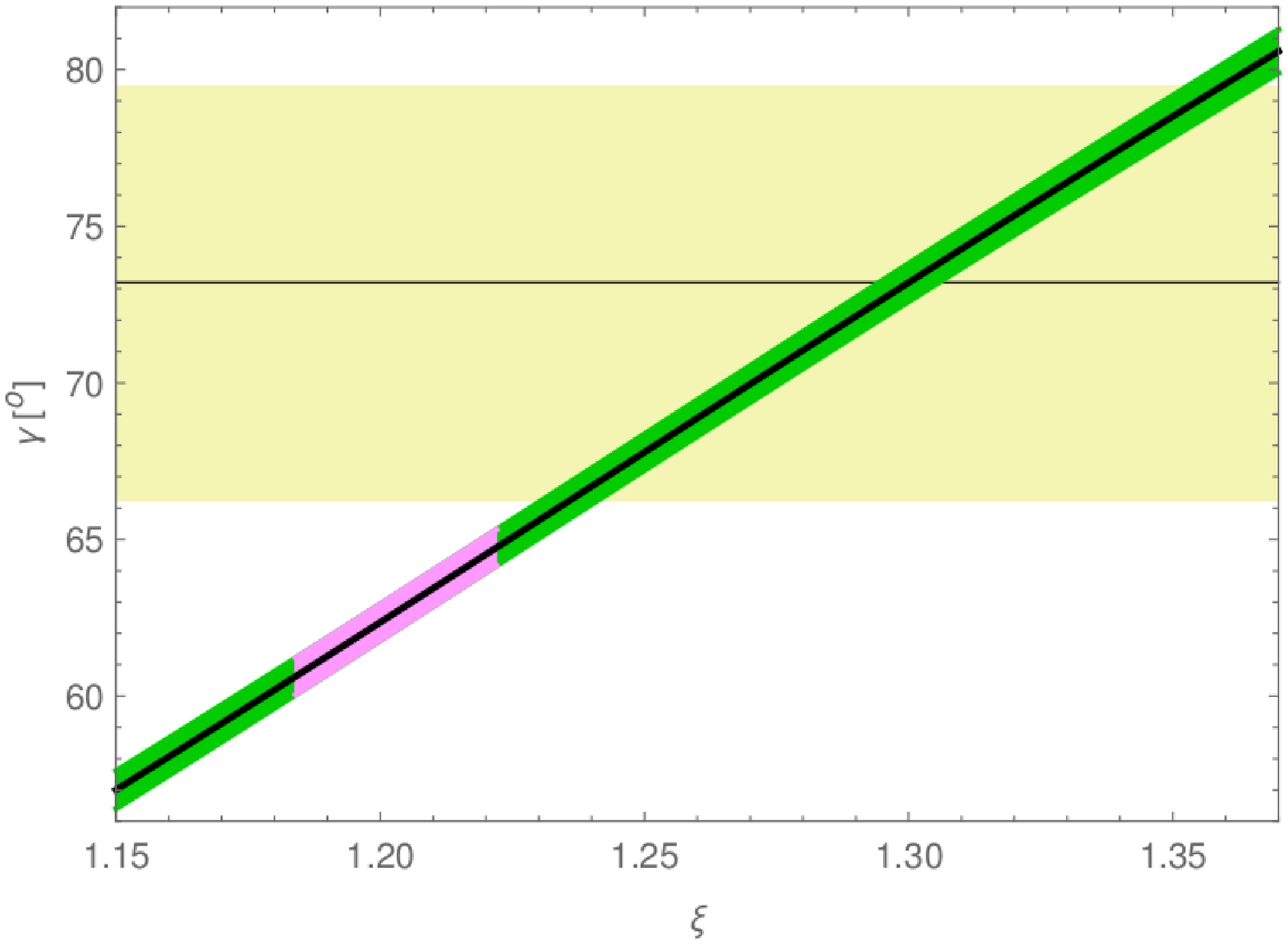} \hfill \includegraphics[width=6.5cm]{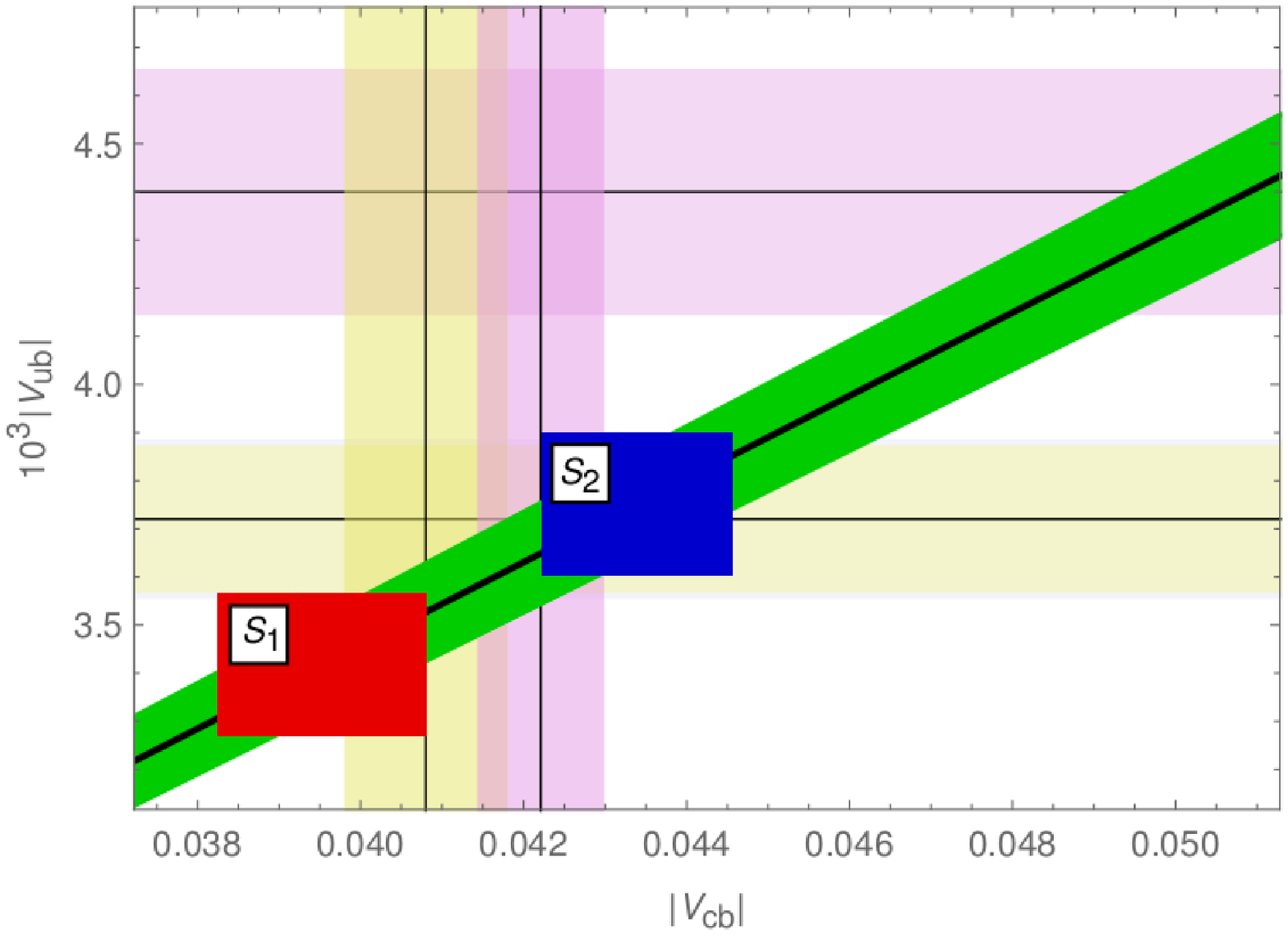} 
\caption{Left: Correlation between $\xi$ and $\gamma$ in CMFV models. The value for $\xi$ from \cite{Bazavov:2016nty} is highlighted in purple. Right: Correlation between $|V_{cb}|$ and $|V_{ub}|$ in CMFV models. From \cite{Blanke:2016bhf}.\label{fig0203}}
\end{figure}

The flavour universality of the function $S(v)$ in \eqref{eq:S} allows to construct the universal unitarity triangle (UUT) \cite{Buras:2000dm} shown in fig.\ \ref{fig01}, which can then be compared with the CKM matrix obtained from tree-level decays. As no new sources of CP-violation are present in CMFV, the angle $\beta$ can directly be obtained from the time-dependent CP-asymmetry
\be
S_{\psi K_S} = \sin2\beta\,.
\ee
The length of the side $R_t$ is determined by the ratio $\Delta M_d/\Delta M_s$ and is therefore very sensitive to the value of $\xi$. In turn, also the angle $\gamma$ exhibits a strong $\xi$ dependence, which is shown in the left panel of fig.\ \ref{fig0203}. The tree-level determination of $\gamma$ \cite{Trabelsi:2014} is depicted by the yellow band. We observe that the rather low value of $\xi$ in \eqref{xi}, highlighted in purple, leads to a surprisingly low value
\be\label{gaUUT}
\gamma_\text{UUT} = (62.7\pm 2.1)^\circ\,.
\ee
Due to the large uncertainty in the tree-level value, the tension is not yet significant. This may however change quickly with the improved precision expected from future LHCb and Belle II measurements. Note that this determination of $\gamma$ does not hold in the more general formulation of MFV where new operators are allowed \cite{D'Ambrosio:2002ex}.

The length $R_b$ of the UUT fixes the ratio 
\be
\left|\frac{V_{ub}}{V_{cb}}\right|_\text{UUT}
=0.0864\pm0.0025\,,
\ee
as shown by the green band in the right panel of fig.\ \ref{fig0203}. Comparing this result with the tree-level determinations of both CKM elements, we notice that the inclusive value of $|V_{ub}|$ is inconsistent with the CMFV hypothesis.

\section{\boldmath Tension between $\Delta M_s$ and $\varepsilon_K$}

In order to fully determine the CKM matrix, one additional experimental input is needed. To this end we employ the following two strategies, with the aim to compare the obtained results:
\begin{itemize}
\item[{\bf\boldmath $S_1$:}]
{\boldmath $\Delta M_s$ \unboldmath} {\bf strategy}  in which the experimental value of 
$\Delta M_s$ is used to determine $\vcb$ as a function of $S(v)$, and $\varepsilon_K$ is then a derived quantity. 
\item[{\bf\boldmath $S_2$:}]
{\boldmath $\varepsilon_K$ \unboldmath}{\bf strategy} in which the experimental value of 
$\varepsilon_K$ is used, while $\Delta M_{s}$ is a derived quantity. 
\end{itemize}
While, within the SM, these strategies entirely fix $|V_{cb}|$ and therefore the entire CKM matrix, in CMFV models $|V_{cb}|$ is determined as a function of the free parameter $S(v)$.

\begin{figure}
\centering{\includegraphics[width=6.5cm]{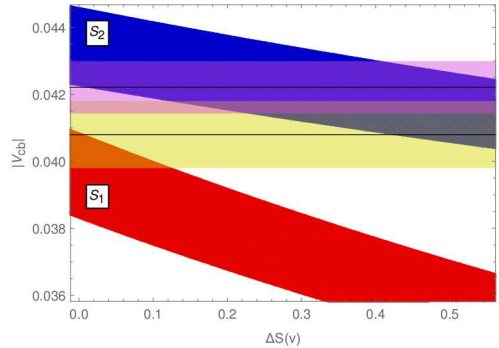}  }   
\caption{$|V_{cb}|$ as a function of $\Delta S(v)$ in scenarios $S_1$ (red) and $S_2$ (blue). From \cite{Blanke:2016bhf}.\label{fig04}}
\end{figure}

The outcome of this exercise is shown in fig.\ \ref{fig04}. We observe that the measured value of $\Delta M_s$ generally requires lower $|V_{cb}|$ values than what is required to account for the data on $\varepsilon_K$. Furthermore, $|V_{cb}|$ in $S_1$ and $S_2$ exhibits a different $S(v)$ dependence. Analytically we find
\bea
\label{S1}
|V_{cb}|_{S_1} &=& (39.5 \pm 1.3)\cdot 10^{-3} \left[\frac{S_0(x_t)}{S(v)}\right]^{1/2} \,,\\
\label{S2}
|V_{cb}|_{S_2} &=& (43.4 \pm 1.2)\cdot 10^{-3} \left[\frac{S_0(x_t)}{S(v)}\right]^{1/4} \,.
\eea
Due to the bound $S(v)\ge S_0(x_t)$ \cite{Blanke:2006yh} these two results cannot be brought in agreement with each other. The tension is in fact smallest in the SM limit.

\begin{figure}
\centering{\includegraphics[width=6.5cm]{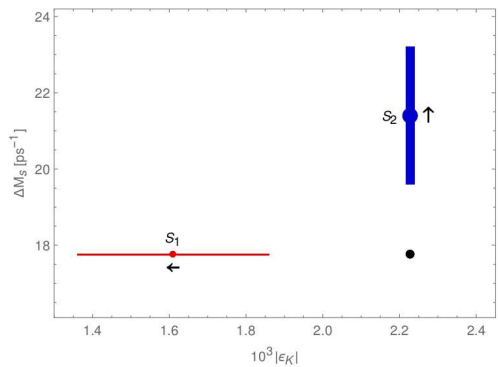}  }   
\caption{$\varepsilon_K$ and $\Delta M_s$ obtained in scenarios $S_1$ (red) and $S_2$ in the SM limit. The arrows indicate how the results change when moving away from $S(v) = S_0(x_t)$. From \cite{Blanke:2016bhf}.\label{fig05}}
\end{figure}

Another way to visualise the tension between $\Delta M_s$ and $\varepsilon_K$ is given in fig.\ \ref{fig05}. In scenario $S_1$, where $\Delta M_s$ is fixed to its experimental value, the prediction for $\varepsilon_K$ lies significantly below the data. Moving away from the SM limit in this case further reduces $\varepsilon_K$ and therefore increases the tension. If instead $\epsilon_K$ is fixed to its experimental value, as done in scenario $S_2$, $\Delta M_s$ turns out above the data and increases with an increasing non-SM contribution to $S(v)$. The same pattern can be found when plotting $\Delta M_d$ versus $\varepsilon_K$. This is not surprising, as the ratio $\Delta M_d/\Delta M_s$ is fixed to its experimental value in the process of constructing the UUT.

Having determined $|V_{cb}|$ in scenario $S_1$ or $S_2$, we can continue to calculate other elements of the CKM matrix. Again, they are functions of $S(v)$, bounded from above by the SM case. Numerical results are collected in table \ref{tab:CKM}. Again we notice that scenarios $S_1$ and $S_2$ yield quite different results.

\begin{table}[!h]
\caption{Upper bounds on CKM elements in units of $10^{-3}$ obtained using the strategies 
$S_1$ and $S_2$. From \cite{Blanke:2016bhf}. 
}\label{tab:CKM}
\begin{tabular}{ccccccc}
\hline
 & $\vts$ & $\vtd$ & $\vcb$ & $\vub$        \\
\hline
$S_1$ &$38.9(13)$   & $7.95(29)$ & $39.5(1.3)$& $3.41(15)$ \\
$S_2$ & $42.7(12)$  &$8.74(27) $ & $43.4(1.2) $& $3.75(15)$ \\
 \hline
\end{tabular}
\end{table}

Before moving on, we use the SM limit summarised in table \ref{tab:CKM} to calulate the SM predictions for some of the cleanest rare decay branching ratios. The result can be found in table \ref{tab:rare}. The difference between scenarios $S_1$ and $S_2$ is quite impressive also in this case.

\begin{table}[!h]
\caption{SM predictions for various rare decay branching ratios using the strategies 
$S_1$ and $S_2$. From~\cite{Blanke:2016bhf}.
}\label{tab:rare}
\begin{tabular}{ccccc}
\hline
 & $ {\mathcal{B}}(\kpn)$ & $ {\mathcal{B}}(\klpn)$ & $\overline{\mathcal{B}}(B_{s}\to\mu^+\mu^-)$ & $\mathcal{B}(B_{d}\to\mu^+\mu^-)$\\
\hline
$S_1$ &$6.88(70)\cdot 10^{-11}$   & $2.11(25)\cdot 10^{-11}$  &$3.14(22)\cdot 10^{-9}$ &$0.84(7)\cdot 10^{-10}$ \\
$S_2$ &$8.96(79)\cdot 10^{-11}$    &$3.08(32)\cdot 10^{-11}$   &$3.78(23)\cdot 10^{-9}$ & $1.02(8)\cdot 10^{-10}$\\
 \hline
\end{tabular}
\end{table}

\section{Going beyond CMFV}

Having identified the problems of CMFV models with $\Delta F =2$ data, implied by the recent Fermilab-MILC results \cite{Bazavov:2016nty}, the question arises how this tension could be resolved. 

A first possibility is, a priori, to relax the lower bound $\Delta S >0$. It has been shown in \cite{Blanke:2006yh} that a negative NP contribution to $S(v)$ is indeed possible. Such a scenario however appears to be rather contrived and is difficult to realise in a concrete NP model. In addition, bringing \eqref{S1} and \eqref{S2} in agreement by means of a negative NP contribution, $S(v) < S_0(x_t)$, results in a prediction for $|V_{cb}|$ significantly above its tree-level determination. Such a scenario is thus not only disfavoured from the model-building perspective, but also by experimental data. 

We therefore conclude that resolving the observed tension requires a flavour non-universal NP contribution to $\Delta F=2$ observables. In general, flavour non-universal NP effects in $\Delta F =2$ processes can be parameterised by replacing the SM loop function $S_0(x_t)$ by the flavour-dependent complex functions
\be
S_i = |S_i| e^{i\varphi_i}\qquad (i=K,d,s)\,.
\ee
These three functions introduce six new parameters in the $\Delta F =2$ sector. It is therefore always possible to fit the available $\Delta F=2$ data and bring them in agreement with the CKM elements determined from tree-level decays. In concrete models, such as the Littlest Higgs model with T-parity \cite{Blanke:2006sb,Blanke:2015wba} or 331 models \cite{Buras:2015kwd}, there exist however correlations between $\Delta F=2$ observables and $\Delta F =1$ rare $K$ and $B$ decays. A more detailed analysis is therefore required. Such a study has recently been presented in the context of 331 models \cite{Buras:2016dxz}.

However, not in all non-CMFV models, all six parameters of the functions $S_i$ are independent. For instance, models with a minimally broken $U(2)^3$ flavour symmetry \cite{Barbieri:2011ci,Barbieri:2012uh}  predict the relations \cite{Buras:2012sd}
\bea
S_K = r_K S_0(x_t) &&\qquad \text{with } r_K > 1\,,\\
S_d=S_s = r_B S_0(x_t) e^{i\varphi_B} &&\qquad \text{with } r_B >0 \text{ and } 0<\varphi_B<2\pi\,.
\eea
A number of interesting predictions arise from this structure. First of all, due to $r_K > 1$ $\varepsilon_K$ can only be enhanced with respect to its SM value. Second, as NP effects are flavour-universal in $B_d-\bar B_d$ and $B_s-\bar B_s$ mixing, they cancel in the ratio $\Delta M_d/\Delta M_s$. The determination of $\gamma$ in \eqref{gaUUT} therefore still holds in $U(2)^3$ models. This may turn out to be problematic if future more precise measurements confirm the tree-level value of $\gamma$ in the ballpark of $70^\circ$.

The relation between the CKM angle $\beta$ and the CP-asymmetry $S_{\psi K_S}$ however is altered due to the presence of the new phase $\varphi_B$.  The ratio $|V_{ub}|/|V_{cb}|$ therefore has to be determined from tree-level decays. Alternatively one can determine $\varphi_B$ from the measured value of $\phi_s$, the CP-violating phase in $B_s-\bar B_s$ mixing, thanks to the smallness of the SM phase $\beta_s$. Subsequently one can use $\varphi_B$ and the measured value of $S_{\psi K_S}$ to determine $\beta$. The triple correlation between $S_{\psi K_S}$, $\phi_s$ and $|V_{ub}|/|V_{cb}|$ thus provides an important consistency check of $U(2)^3$ models.

\section{Summary and outlook}

The improved lattice results for the hadronic matrix elements entering $B_{d,s}-\bar B_{d,s}$ mixing presented recently by the Fermilab-MILC collaboration \cite{Bazavov:2016nty} imply a tension in $\Delta F=2$ data not only within the SM, but also more generally in CMFV models \cite{Blanke:2016bhf}. 

At present the situation is still unconclusive. On the one hand, the Fermilab-MILC results need to be confirmed (or challenged) by other lattice collaborations before we can settle on the values of the hadronic matrix elements in question. On the other hand, more precise experimental data are required, in particular on the tree-level determinations of the CKM elements $\vub$, $\vcb$ and $\gamma$, but also on the CP-violating observables $S_{\psi K_S}$ and $\phi_s$. Last but not least,  a further reduction of uncertainties in $\varepsilon_K$ would be very desirable in order to disentangle the NP flavour structure.

\acknowledgments

My thanks go to Andrzej J.\ Buras for the pleasant and fruitful collaboration which led to the results presented here.  
I would also like to thank the organisers of the XXX Rencontres de Physique de la Vall\'ee d'Aoste for inviting me to La Thuile and giving me the opportunity to present these results.

\end{document}